\documentclass[twocolumn]{aastex701}

\usepackage{subfigure}
\usepackage{graphicx}
\usepackage{amsmath}
\usepackage{enumitem}

\begin{document}

\title{The disk precession in a Be star-magnetar binary and its application to the rotation measure of FRB 20201124A}

\author{Ying-ze Shan}
\affiliation{Department of Astronomy, School of Physics, Huazhong University of Science and Technology, Wuhan, 430074, China}
\email{}  

\author[orcid=0000-0003-3440-1526]{Wei-Hua Lei} 
\affiliation{Department of Astronomy, School of Physics, Huazhong University of Science and Technology, Wuhan, 430074, China}
\email[show]{leiwh@hust.edu.cn}

\author{Hao-Tian Lan}
\affiliation{School of Astronomy and Space Science, Nanjing University, Nanjing 210093, China}
\email{}

\author{Shao-yu Fu}
\affiliation{Department of Astronomy, School of Physics, Huazhong University of Science and Technology, Wuhan, 430074, China}
\email{}

\author{Jumpei Takata}
\affiliation{Department of Astronomy, School of Physics, Huazhong University of Science and Technology, Wuhan, 430074, China}
\email{}

\author[orcid=0000-0002-5400-3261]{Yuan-chuan Zou}
\affiliation{Department of Astronomy, School of Physics, Huazhong University of Science and Technology, Wuhan, 430074, China}
\email[show]{zouyc@hust.edu.cn}

\author{Jia-xin Liu}
\affiliation{Department of Astronomy, School of Physics, Huazhong University of Science and Technology, Wuhan, 430074, China}
\email{}

\author{Long-xuan Zhang}
\affiliation{Department of Astronomy, School of Physics, Huazhong University of Science and Technology, Wuhan, 430074, China}
\email{}

\author{Tong-lun Wang}
\affiliation{Department of Astronomy, School of Physics, Huazhong University of Science and Technology, Wuhan, 430074, China}
\email{}

\author[orcid=0000-0003-4157-7714]{Fa-Yin Wang} 
\affiliation{School of Astronomy and Space Science, Nanjing University, Nanjing 210093, China}
\affiliation{Key Laboratory of Modern Astronomy and Astrophysics (Nanjing University), Ministry of Education, Nanjing 210093, China}
\email[show]{fayinwang@nju.edu.cn}

\begin{abstract}

Fast radio bursts (FRBs) are bright, millisecond-duration radio bursts with poorly known origins. Most FRB sources are detected only once, while some are repeaters. Variation patterns observed in the rotation measure (RM) of some repeaters indicate that the local magneto-ionic environments of these FRB sources are highly dynamic. It has been suggested that a Be star-magnetar binary system is a possible origin for such variation. FRB 20201124A is notable among these sources since it is one of the very active repeaters and exhibits substantial temporal variations of RM measured by the Five-hundred-meter Aperture Spherical radio Telescope (FAST). The physics behind this long-term behavior remains poorly understood. Here we propose that, within the framework of the Be star-magnetar binary scenario, the observed variation in RM is attributed to a combination of orbital motion and the precession of the circumstellar disk of the Be star. While a $
\sim785$-day precession of the disk contributes to the observed decrease in the amplitude of the variation, our model predicts that the amplitude oscillates with this period.                                                            
\end{abstract}

\section{Introduction} \label{1}
FRBs are a class of cosmological radio transients with millisecond durations and high luminosity \citep{Lorimer_2007Sci...318..777L, Thornton_2013Sci...341...53T, Wu2024}. Although the origin of FRBs remains poorly understood, the detection of FRB 200428 from the Galactic magnetar SGR J1935+2154 reveals that magnetars are a viable source of FRBs \citep{Bochenek_2020Natur.587...59B, CHIME_2020Natur.587...54C}. Many researchers indicate that FRB emission regions are likely to be within magnetar magnetospheres \citep{Kumar_2017MNRAS.468.2726K, Yang2018, Nimmo_2025Natur.637...48N}. Approximately 4,000 FRB sources have been detected to date \citep{2026arXiv260109399F}, most of which are detected only once, whereas others exhibit repeated emissions \citep{Spitler_2016Natur.531..202S, CHIME_2019ApJ...885L..24C,  CHIME_2023ApJ...947...83C}. Long-term observations of FRB repeaters allow for the investigation of their properties, e.g., dispersion measure (DM), rotation measure (RM), polarization, burst rate, and waiting time, helping to reveal the origin of FRBs.

Over a decade of observations of FRBs, several FRBs have exhibited dramatic RM variations and oscillations \citep{Hilmarsson_2021MNRAS.508.5354H, Xu_2022Natur.609..685X, Kumar_2023MNRAS.526.3652K, Anna-Thomas_2023Sci...380..599A, Mckinven_2023ApJ...950...12M, Ng_2025ApJ...982..154N}. Since the RM is related to the magnetic field and plasma density, the temporal variation may indicate the evolution of the local magneto-ionic environment around the FRB source. 

Among these FRBs, FRB 20201124A attracts significant attention. It is an extremely active repeater and exhibits peculiar variations in its RM. From the observations by FAST between April and June 2021 \citep{Xu_2022Natur.609..685X}, the RM of FRB 20201124A exhibits a pattern of rapid increase, sharp drop, and subsequent re-ascent over $\sim$ 40 days, with the maximum RM variation reaching $\sim$ 500 $\rm{rad/m^2}$. The variation stage is overlaid by a series of oscillations, and followed by a stable stage lasting for $\sim$ 20 days. After subtracting the RM averaged over $\sim$ two-month observations (from 1 April to 11 June, 2021), the variations show that the local RM contribution of FRB 20201124A changes its sign \citep{Wang_2022NatCo..13.4382W}. 

To explain the origin of repeating FRBs, a model involving the interaction between a pulsar magnetosphere and its companion star's stream, i.e., the "cosmic-comb" model, has been proposed by \cite{Zhang_2017ApJ...836L..32Z}. Later this model is adopted in \cite{Zhang_2018ApJ...854L..21Z} to explain the observational properties of FRB 20121102. After the 16-day periodically modulated activity for FRB 20180916B is found, more studies focused on the relation between FRBs and the orbital motion of binary systems are conducted \citep{Ioka_&_Zhang_2020ApJ...893L..26I, Katz_2020MNRAS.494L..64K, Sridhar_2021ApJ...917...13S, Pleunis_2021ApJ...911L...3P} \footnote{These models cannot account for the observed frequency-dependent active window of FRB 20180916B \citep{2021Natur.596..505P}, until \cite{Wada_&_Ioka_&_Zhang_2021ApJ...920...54W} offers an explanation via the ``cosmic-comb'' model.}.

For the observed RM variation of FRB 20201124A, \cite{Wang_2022NatCo..13.4382W} examined a Be star and magnetar binary scenario and suggested that the orbital motion causes a variation in the local magneto-ionic environment along the line of sight (LoS), resulting in the sign-changing RM. This model is supported by the following studies \citep{Xu_2025arXiv250506006X}. \cite{Zhao_&_Zhang_&_Wang_&_Dai_2023ApJ...942..102Z} points out that the RM variation of FRB 20180916B may be caused by the stellar wind of a massive companion. \cite{Xu_2025arXiv250506006X} claimed the detection of a $\sim$ 26-day periodicity with significance $>5\sigma$ in the RM observations of FRB 20201124A from 2021 to 2022. \cite{Du_2025arXiv250312013D} detected a $\sim$ 1.7 s periodicity with high significance in the times of arrival (ToAs) of FRB 20201124A bursts observed on MJD 59310 and MJD 59347; such a period is explained by the spinning of the potential magnetar central engine of FRB 20201124A. \cite{Zhang_2025arXiv250517880Z} further noted that the two days with detected ToAs periodicity bookend the RM variation phase of FRB 20201124A, and they might be the two observational windows when the magnetar central engine crossed the Be star disk. In addition to FRB 20201124A, \cite{Li_doi:10.1126/science.adq3225} reported that the RM of FRB 20220529 shows sudden variations within a month, and \cite{Liang_2025ApJ...994L..32L} reported the existence of a $\sim$ 200-day periodicity in the RM of FRB 20220529 from 2022 to 2025. These potential periodicities in RM are likely to be related to the orbital motion of binary star systems \citep{Rajwade2023,ZhangB2025,Wang2025}. 

Super-orbital brightness variations --- periodic brightness oscillations far longer than the binary’s orbital period in multi-wavelength bands --- have been observed in many Be star-neutron star binaries \citep{Ogilvie_2001MNRAS.320..485O, McGowan_2008MNRAS.384..821M, Rajoelimanana_2011MNRAS.413.1600R} --- specifically a Be/X-ray binary \citep{Negueruela_1998A&A...336..251N, Coe_2005MNRAS.356..502C, Liu_2005A&A...442.1135L, Haberl_&_Sturm_2016A&A...586A..81H}. While the driving mechanism of this periodic behavior is unclear, many studies indicate that it could be induced by Be disk precession \citep{Rajoelimanana_2011MNRAS.413.1600R, Martin_&_Franchini_2021ApJ...922L..37M, Martin_2023MNRAS.523L..75M, Martin_&_Charles_2024MNRAS.528L..59M, Chen_2024ApJ...973..162C}. Assuming that the rotation axis of the Be star precesses with a superorbital period, and according to \cite{Martin_2023MNRAS.523L..75M}, the decretion disk can remain locked to the equatorial plane of the Be star; the node of the disk plane and the neutron star orbit will precess with the star. Since the brightness variations of the Be star essentially reflect changes in the disk density along the LoS between the Be star and the observer, the disk precession can lead to the periodic modulation of the density profile along the line of sight, producing the super-orbital modulation in the observed brightness 
\citep{Alcock_2001MNRAS.321..678A, Rajoelimanana_2011MNRAS.413.1600R, Martin_&_Charles_2024MNRAS.528L..59M}. Moreover, X-ray bursts emitted when the neutron star interacts with the disk may also have their brightness modulated by the precession of the Be star and the disk \citep{Alcock_2001MNRAS.321..678A}. The magnitude of FRB 20201124A RM varies significantly in three active epochs, which cannot be explained by simple binary orbital motion. The above studies motivate us to investigate the precession of the Be star disk in binary systems and its effect on modulating the RM variations of FRBs.  

In this work, we present a Be star-magnetar model with a precessing disk to elucidate the long-term RM variations in FRB 20201124A. The paper is organized as follows. The observations and data reduction of FRB 20201124A are presented in Section \ref{2}. The model of a binary system with a precessing disk is described in Section \ref{3}. The fitting results are shown in Section \ref{4}. In Section \ref{5}, we draw conclusions and discuss other FRB sources with similar RM variations.

\section{FRB 20201124A} \label{2}
FRB 20201124A is a highly active repeating FRB source. By April 2021, FRB 20201124A entered an extremely active phase \citep{Xu_2022Natur.609..685X}. In addition, FRB 20201124A is found to be associated with a persistent radio source (PRS), which is consistent with the magnetar origin hypothesis for this FRB \citep{Bruni_2024Natur.632.1014B}. Since 2021, FAST has been monitoring this source, covering the 1.0–1.5 GHz frequency range, and has detected thousands of bursts during the three active cycles \citep{Xu_2022Natur.609..685X,Jiang_2022RAA....22l4003J,Niu_2022RAA....22l4004N,Xu_2025arXiv250506006X}. The three FAST observation epochs are hereinafter referred to as E1, E2, and E3 for brevity. Only the bursts with a high signal-to-noise ratio $S/N$ are adopted to measure the RMs using a Q-U fitting method \citep{Desvignes_2019Sci...365.1013D}.

In E1, FAST monitored FRB 20201124A from April 1, 2021, to June 11, 2021 (UT). During a total observing time of 91 hours, 1,863 bursts were detected. A total of 1,103 bright bursts with a $S/N > 30$ are adopted to measure the RM \citep{Xu_2022Natur.609..685X}. Intriguingly, the RM of the adopted bursts shows dramatic variation in a short term ($\sim 60$ days), and the maximum variation can reach about 500 $\rm{rad\,m^{-2}}$, making it the first FRB observed with rapid RM variation in such a short timescale \citep{Xu_2022Natur.609..685X,Wang_2022NatCo..13.4382W}.

E2 spans from 25 September 2021 to 17 October 2021 (UT), with a total observation of 19 hours. However, a total of 881 bursts was observed during the first four days, i.e., 25-28 September 2021, whereas no bursts were detected later in this epoch \citep{Zhou_2022RAA....22l4001Z,Zhang_2022RAA....22l4002Z,Jiang_2022RAA....22l4003J,Niu_2022RAA....22l4004N}. Among these 881 bursts, 536 bursts with $S/N \geq 50$ are adopted to measure RM. The RM of the E2 bursts remains at the average level of the E1 bursts, but displays a feature of continuous drop. swinging and orthogonal jumping PAs are also found in some of these bursts \citep{Niu_2024ApJ...972L..20N}.

E3 was conducted by FAST from 2 February 2022 to 26 May 2022, during the third active cycle. In this epoch, 1,467 bright bursts with $S/N>20$ are detected in a total observing time of 34.3 hours \citep{Xu_2025arXiv250506006X}. The RM data of the E3 bursts also show a variation feature within several tens of days, but at a level of $\sim100\,\rm{rad\,m^{-2}}$, which is a few times lower than that of epoch one. \cite{Xu_2025arXiv250506006X} searched for the periodicity of the RM variation in the E3 bursts, and a period of $\sim30$ days was found with high significance. We note that the E3 RM data have a lower background level compared to that of E1 and E2, which might be due to the long-term evolution of the magneto-ionic environment. After E3, this source falls into quiescence.

In this work, we use the FRB 20201124A RM data of E1, E2, and E3 (denoted as $\rm{RM_{E1}}$, $\rm{RM_{E2}}$, and $\rm{RM_{E3}}$ below, respectively). Data of $\rm{RM_{E1}}$ are provided by \citep{Xu_2022Natur.609..685X}, and data of $\rm{RM_{E2}}$ and $\rm{RM_{E3}}$ are obtained from the plots in \cite{Xu_2025arXiv250506006X} using the online data extraction tool {\sc{WebPlotDigitizer}} \citep{Marin_2017sf2a.conf..113M}. To characterize RM variations, we perform some pre-processing operations on $\rm{RM_{E1}}$, $\rm{RM_{E2}}$, and $\rm{RM_{E3}}$. For each episode (episode $i$), we first bin $\rm{RM_{E\textit{i}}}$ by MJD and derive the mean RM value in each MJD bin (after this step, the data of episode $i$ are then denoted as $\rm{RM_{dE\textit{i}}}$). Secondly, we derive the mean for $\rm{RM_{dE\textit{i}}}$, which gives $\overline{\rm{RM}}_{\rm{dE1}}=-583.95\,\rm{rad}\,m^{-2}$, $\overline{\rm{RM}}_{\rm{dE2}}=-593.35\,\rm{rad}\,m^{-2}$, and $\overline{\rm{RM}}_{\rm{dE3}}=-666.45\,\rm{rad}\,m^{-2}$. Finally, we can obtain the RM variations $\Delta \rm{RM}_{E1}$, $\Delta \rm{RM}_{E2}$, and $\Delta \rm{RM}_{E3}$ by calculating $\Delta \rm{RM}_{E\textit{i}}=\rm{RM_{dE\textit{i}}}-\overline{\rm{RM}}_{dE\textit{i}}$.

\section{Be star-magnetar binary with disk precession} \label{3}

\subsection{Be star-magnetar binary} \label{3.1}

We consider a Be star-magnetar binary system. Stellar outflow from the Be star can be particularly significant in the equatorial region, where dense flow usually forms a decretion disk. The density of the decretion disk decreases with increasing radius, and the base density $\rho_{0}$ typically ranges from about $10^{-12}$ to $10^{-10}\,\rm{g\,cm^{-3}}$ \citep{Porter_1999A&A...348..512P,Gies_2007ApJ...654..527G,Rivinius_2013A&ARv..21...69R,Wang_2022NatCo..13.4382W}. 

Here, we consider an axisymmetric disk and describe the density profile as
\citep{Wang_2022NatCo..13.4382W}
\begin{equation}
    \rho(r,z)=\rho_0(\frac{r}{R_{\star}})^{-\beta}{\rm{exp}}[-(\frac{D}{H(r)})^2],
    \label{(1)}
\end{equation}
\noindent where $\rho_0$, $r$, $D$, $R_\star$, and $\beta$ are the disk density at the stellar radius, the distances from the stellar center, the heights from the stellar equatorial plane, the stellar radius, and the density slope index, respectively. In the case of a Be star, $\beta$ usually ranges from $2\sim4$. In addition, $H(r)$ is the vertical scale height of the disk, which is determined by  \citep{Bjorkman_1997LNP...497..239B,Rivinius_2013A&ARv..21...69R}
\begin{equation}
    H(r)\simeq c_{\rm s}(\frac{r}{GM_\star})^{1/2}r，
\label{(2)}
\end{equation}
\noindent where $G$ is the gravitational constant, $M_\star$ is the stellar mass, and $c_{\rm s}$ is the sound speed in the disk. For an isothermal gas disk, $c_{\rm s}$ can be given by
\begin{equation}
    c_{\rm s}=\sqrt{\frac{kT}{\mu m_{\rm H}}},
\label{(3)}
\end{equation}
\noindent where $k$, $T$, $\mu$, and $m_{\rm H}$ are the Boltzmann constant, the isothermal temperature, the mean molecular weight, and the atomic mass of hydrogen, respectively. 

Previous studies indicate a magnetic field is present in the outflow disk of a early type star like Be star \citep{Usov_&_Melrose_1992ApJ...395..575U,Melatos_&_Johnston_&_Melrose_1995MNRAS.275..381M, Wang_2022NatCo..13.4382W, Zhao_&_Zhang_&_Wang_&_Dai_2023ApJ...942..102Z}. Here, we adopt a simplified toroidal magnetic field at a relatively large distance from the Be star: 
\begin{equation}
    \textbf{B}(\textbf{r})={\rm{B_0}}(\frac{R_\star}{r})\vec{e}_{r}\times \vec{e}_{\rm{disk}},
\label{(4)}
\end{equation}
\noindent where $\rm{B_0}$, \textbf{r}, $\vec{e}_{r}$, and $\vec{e}_{\rm{disk}}$ are the initial strength of the toroidal magnetic field, the vector pointing from the stellar center to the field point, the unit vector of \textbf{r}, and the unit normal vector of the Be star disk, respectively. A more detailed discussion on the magnetic field configuration of Be stars can be seen in Section \ref{5.1.4}.

The magnetar orbits the Be star along an elliptical path following Keplerian motion. The semi-major axis of the orbit can be expressed as
\begin{equation}
    a=\left[\frac{G(M_\star +M_{\rm{NS}})P_{\rm{orb}}^2}{4\pi^2}\right]^{1/3},
\label{(5)}
\end{equation}
\noindent where $M_{\rm{NS}}$ is the magnetar mass, and $P_{\rm{orb}}$ is the orbital period of the magnetar. The mass of the Be star is about 10 $M_\odot$, while the NS mass ranges from 1$\sim$2 $M_\odot$, which is much lower than that of the Be star. Thus, we can assume that the Be star is approximately located at one focus of the elliptical orbit. Be-neutron star binary systems are often observed with high orbital eccentricity, which is likely caused by the natal kick of the supernova explosion in the binary system.

We assume that the magnetar continuously emits FRB in its magnetosphere while moving along its orbit. The FRB propagating along the observer's LoS may pass through the outflow material of the star, mainly the decretion disk. Thus, the density profile and magnetic field along the LoS in the disk determine the observed RM:
\begin{equation}
   {\rm{RM}} = 8.1\times 10^5\int_{\rm{LoS}} \frac{n_{\rm e}(l)\cdot\textbf{B}_\parallel(l)}{(1+z(l))^2}\,\textit{dl}\,\,\,\rm{rad\,m^{-2}},
\label{(6)}
\end{equation}
\noindent where $\textbf{B}_\parallel$ is the magnetic field component along the LoS (in units of Gauss), $z$ is the redshift, and $n_{\rm e}$ is the number density of free electrons (in units of $\rm{cm^{-3}}$).
Assuming that the disk is mainly composed of hydrogen, the number density of the free electrons for the gas temperature $T$ is calculated from the Saha ionization equation:
\begin{equation}
   \frac{n_{\rm H^+} \cdot n_{\rm e}}{n_{\rm H}} = \frac{2g_{\rm H^+}}{g_{\rm H}} \left( \frac{2\pi m_{\rm e} k T}{h^2} \right)^{3/2} e^{-\chi_0 / (k T)},
\label{(7)}
\end{equation}
\noindent where $g_{\rm H^+}$ and $g_{\rm H}$ are the statistical weights of the proton and neutral hydrogen atom, respectively (we set $g_{\rm H^+}=g_{\rm H}$ for simplification), $n_{\rm H^+}$ is the number density of ionized hydrogen atoms, and we assume that $n_{\rm H^+}=n_{\rm e}$ according to the conservation of charge. In addition, $m_{\rm e}$, $h$, and $\chi_0=13.6\,\rm{eV}$ are the electron mass, Planck constant, and ionization energy of the hydrogen atom, respectively.

\begin{figure*}
    \vspace{1cm}
 	\centering
 	\includegraphics[width=1\textwidth, height=0.8\textheight, keepaspectratio]{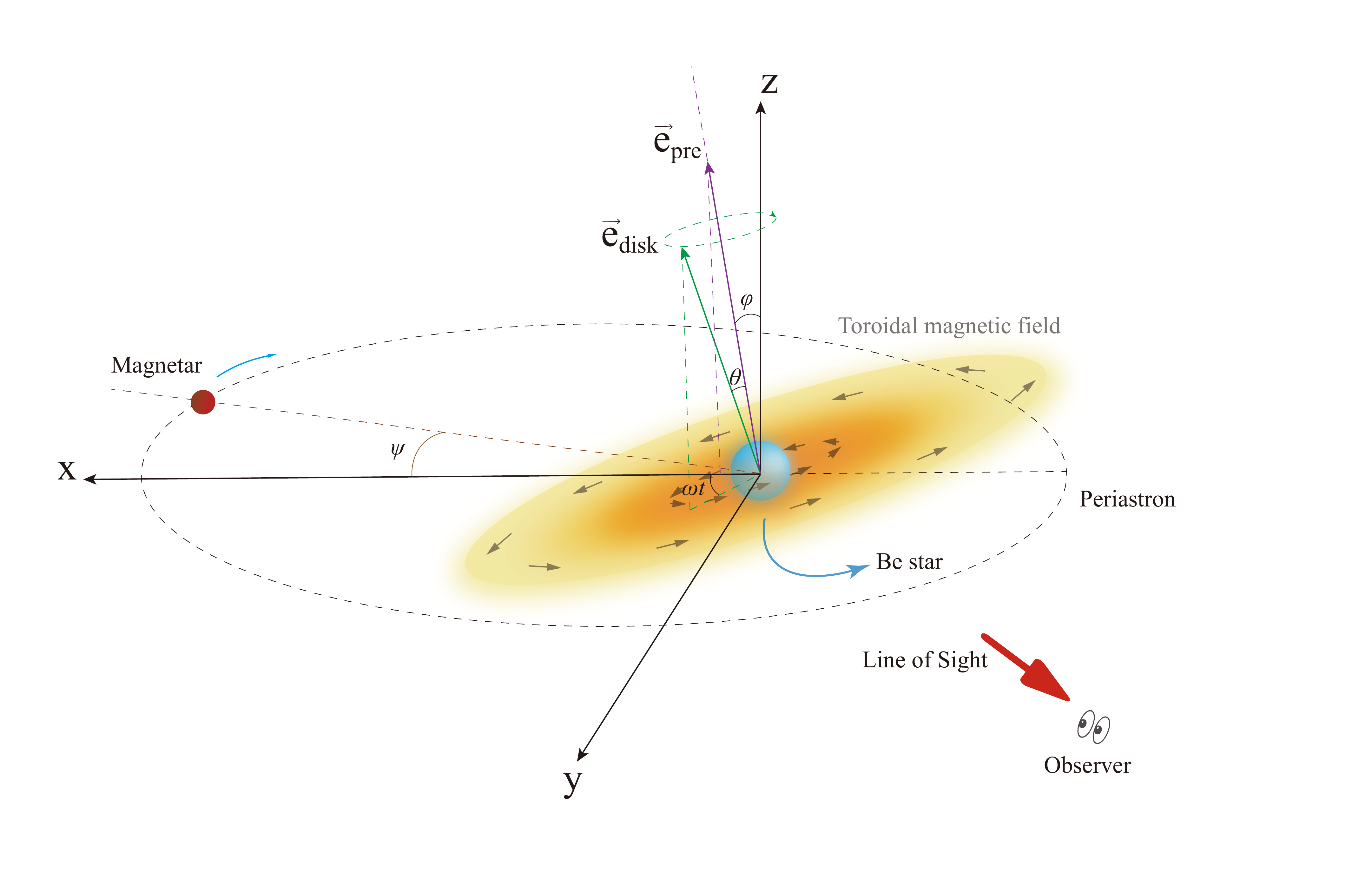}
\caption{A schematic diagram of the binary system in Section \ref{3} (not to scale). Note that the darker regions in the disk have higher density. The magnetar continuously emits FRBs and orbits the Be star along an elliptical orbit. FRBs emitted by the magnetar at different positions along its orbit propagate along the LoS and reach the observer. Precession of the Be star disk leads to varying RMs among different bursts.} \label{fig:1}
\end{figure*}

\subsection{Disk precession} \label{3.2}
In a Be star-NS binary, the supernova natal kick at the birth of the NS may lead to a misalignment of the orbital plane and the Be star disk plane \citep{Brandt_1995MNRAS.274..461B},as well as a binary orbit with high eccentricity \citep{van_den_Heuvel_1997ApJ...483..399V}.

Following the method of \cite{Chen_2024ApJ...973..162C}, we set a Cartesian coordinate system where the origin is at the center of the star, where the $x$-axis points toward the periastron of the binary star orbit, and the $z$-axis is perpendicular to the orbital plane (see Figure \ref{fig:1}). The unit vectors of the $x$-, $y$- and $z$-axes, are denoted by $\vec{e}_{x}$, $\vec{e}_{y}$ and $\vec{e}_{z}$, respectively. The unit direction vector of the magnetar $\vec{e}_{\rm{mag}}$, and the angle between it and $x$ axis is $\psi$. The relation between the true anomaly $\psi$ and the time $t$ (orbital phase) is described by Kepler's equation:
\begin{equation}
    \frac{2\pi t}{P_{\rm{orb}}}=2\rm{arctan}(\sqrt{\frac{1-\varepsilon}{1+\varepsilon}}\rm{tan}(-\psi/2))+\varepsilon\frac{\rm{sin}\psi \sqrt{1-\varepsilon^2}}{1+\varepsilon \rm{cos}\psi},
\label{(8)}
\end{equation}
\noindent where $\varepsilon$ is the eccentricity of the magnetar orbit.

The unit vector pointing from the Be star to the center of the magnetar can be expressed using the rotation matrix as follows:
\begin{equation}
    \vec{e}_{\rm{mag}}=\textbf{R}[\vec{e}_{z}, -\psi]\vec{e}_{x}=
    \begin{bmatrix}
    \rm{cos\psi}\\
    \rm{-sin\psi}\\
0
\end{bmatrix}.
\label{(9)}
\end{equation}

The Be star disk precesses around a precession axis, which can be defined by a unit vector $\vec{e}_{\rm{pre}}$ and forms an angle $\varphi$ with the $z$ axis. To simplify the scenario, $\varphi$ is constrained to the $x$-$z$ plane. $\theta$ is the angle between $\vec{e}_{\rm{disk}}$ and $\vec{e}_{\rm{pre}}$. If the period of precession is $P_{\rm{pre}}$, $\vec{e}_{\rm{pre}}$ and $\vec{e}_{\rm{disk}}$ can be expressed as follows:
\begin{equation}
    \vec{e}_{\rm{pre}}=\textbf{R}[\vec{e}_{y}, -\varphi]\vec{e}_{z}=
\begin{bmatrix}
\rm{-sin\varphi}\\
0\\
\rm{cos\varphi}
\end{bmatrix},
\label{(10)}
\end{equation}

\begin{equation}
    \vec{e}_{\rm{disk}}=\textbf{R}[\vec{e}_{\rm{pre}}, \omega (t-t_0)]\vec{e}_{\rm{disk,0}},
\label{(11)}
\end{equation}
\noindent where $\omega =2\pi /P_{\rm{pre}}$ is the angular velocity of precession, and $\vec{e}_{\rm{disk,0}}$ is the unit vector normal to the disk at the initial time $t_0$. To simplify, we define $\vec{e}_{\rm{disk,0}}$ as follows:
\begin{equation}
    \vec{e}_{\rm{disk,0}}=\textbf{R}[\vec{e}_{y}, -(\frac{\pi}{2}+\theta +\varphi)]\vec{e}_{x}=
\begin{bmatrix}
    \rm{-sin(\theta +\varphi)}\\
    0\\
    \rm{cos(\theta +\varphi)}
\end{bmatrix},
\label{(12)}
\end{equation}
\noindent where $\theta$ is the angle between $\vec{e}_{\rm{disk,0}}$ and $\vec{e}_{\rm{pre}}$.

In this coordinate system, the unit direction vector of LoS can be denoted as 
\begin{equation}
    \vec{e}_{\rm{LoS}}=
\begin{bmatrix}
    \rm{cos\lambda_1 sin\lambda_2}\\
    \rm{sin\lambda_1 sin\lambda_2}\\
    \rm{cos\lambda_2}
\end{bmatrix},
\label{(13)}
\end{equation}
\noindent where $\lambda_1$ or $\lambda_2$ are the angles between $\vec{e}_{\rm{LoS}}$ and $\vec{e}_{x}$ or $\vec{e}_{z}$.

\section{Application to FRB 20201124A} \label{4}

In this section, we will use the precession disk model of the B star-magnetar binary in Section \ref{3.2} to explain the variations in FRB 20201124A RM data. 

\subsection{Model parameters} \label{4.1}
In our model described in Section \ref{3}, the parameters are as follows

\begin{enumerate}[label=., left=0pt, itemindent=1em]
    \item Stellar parameters:
    $M_{\rm{NS}}$, $M_\star$, and $R_\star$; 
    \item Disk properties:$\rm{B_0}$, $\rho_0$, $\mu$, $\beta$, and $T$;
    \item Parameters of the binary orbit: $\varepsilon$ and $P_{\rm{orb}}$;
    \item Parameters of the disk precession: $\varphi$ of $\vec{e}_{\rm{pre}}$, $\theta$ of $\vec{e}_{\rm{disk,0}}$, and $\omega$ of $\vec{e}_{\rm{disk}}$;
    \item LoS: $\lambda_1$ and $\lambda_2$ of $\vec{e}_{\rm{LoS}}$.
\end{enumerate}
 
For stellar parameters, their values are fixed in our calculations. For a B type star, its mass range is poorly constrained, spanning from a handful to a dozen $M_\odot$. In this study, we fix $M\star$ as 8 $M_\odot$. Since the radius of B-type stars ranges from a few to ten $R_\odot$, we adopt $R_\star=5\,R_\odot$. For the magnetar mass $M_{\rm{NS}}$, we use a typical NS mass of 1.6 $M_\odot$. 

For disk properties, the isothermal temperature, the mean molecular weight, and the initial magnetic field in the disk are assumed to be $T=20,000\,\rm{K}$, $\mu=1$, and $\rm{B_0}=10\,\rm{G}$, respectively, referring to \cite{Wang_2022NatCo..13.4382W}. Since the fitting is sensitive to the density profile of the disk, we set $\rho_0$ and $\beta$ as free parameters to achieve a better fitting result. 

\begin{figure*}
    \vspace{1cm}
 	\centering
 	\includegraphics[width=1\textwidth, height=0.8\textheight, keepaspectratio]{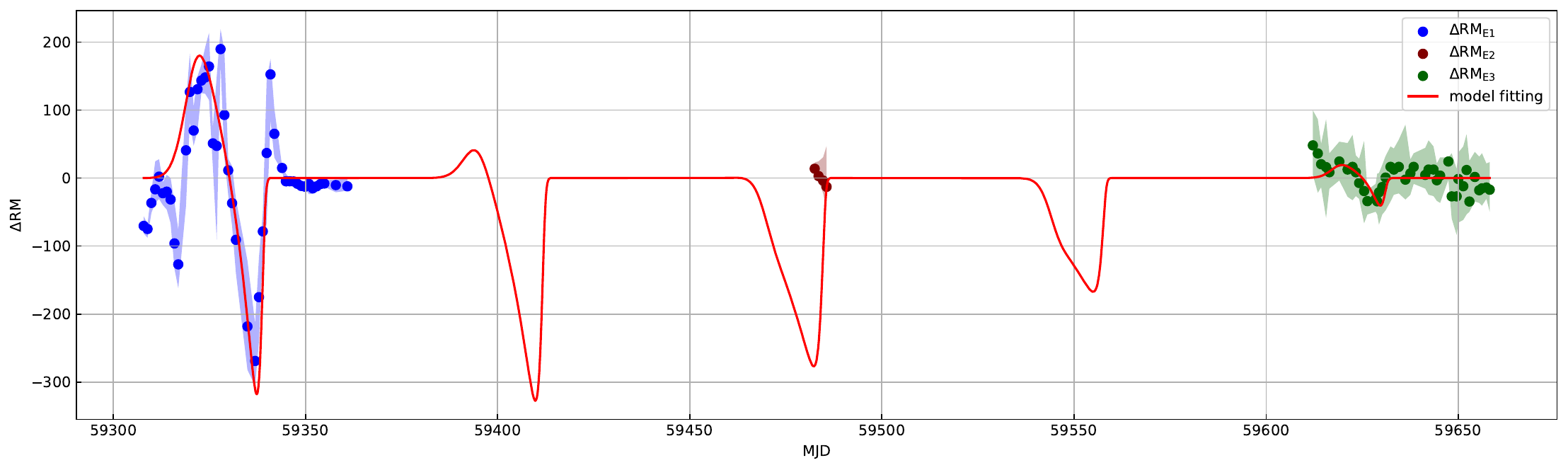}
\caption{The RM variations of FRB 20201124A in three FAST observation episodes, i.e., $\Delta\rm{RM}_{E1}$ (blue dots), $\Delta\rm{RM}_{E2}$ (maroon dots) and $\Delta\rm{RM}_{E3}$ (dark green dots). The observed RM ranges of E1, E2 and E3 are shown as the blue, maroon, and green shadow regions, respectively. The fit of the model presented in Section \ref{3} is shown as the red curve, and the corresponding fitting parameters are listed} in Table \ref{tab:1}. \label{fig:2}
\end{figure*}

For orbital parameters, i.e., $P_{\rm{orb}}$ and $\varepsilon$, we fix $P_{\rm{orb}}$ but treat $\varepsilon$ as a free parameter. According to the RM data of FRB 20201124A and our model, we anticipate that the maximum dips of E1 ($\sim$ MJD 59337) and E3 ($\sim$ MJD 59627) as seen in Figure \ref{fig:2} are very likely in the same phase of magnetar orbital motion. Hence, our model expects that the time span between the two dips is an integer multiple of the orbital period $P_{\rm{orb}}$ and estimates the period to be $P_{\rm{orb}}=73$ days. With $P_{\rm{orb}}=73$ days, four orbital periods have been covered by the current FAST observations (see Figure \ref{fig:2}). Since the orbital eccentricity $\varepsilon$ can significantly affect the profile of the RM variations, it is treated as a free parameter.

For parameters of disk precession and LoS, as they are directly related to precession and RM integration, all of them are set as free parameters. In conclusion, our fitting parameters are $\varepsilon$, $\rho_0$, $\beta$, $\varphi$, $\lambda_1$, $\lambda_2$, $\theta$, and $\omega$.

\subsection{Results} \label{4.2}
To compute the LoS integral of RM, we apply the method developed by \cite{Wang_2022NatCo..13.4382W}: At each orbital phase, we take 1000 sample points along the LoS starting from the magnetar’s position, with a distance of $1\,R_\odot$ between two adjacent points. At each calculation point, the local plasma density and the magnetic field are calculated from Equations (\ref{(1)}) and (\ref{(4)}), respectively, and the expected $\Delta\rm{RM}$ is calculated from Equation (\ref{(6)}). 

The orbital eccentricity $\varepsilon$ mainly affects the ratio between the peak and dip durations, as well as the amplitude of the $\Delta\rm{RM}$ curve. A larger $\varepsilon$ increases the relative duration of the peak within one orbital period and leads to more pronounced variations in the $\Delta\rm{RM}$ curve. The base mass density $\rho_0$ and the density radial profile $\beta$ can also influence the variation amplitude of the $\Delta\rm{RM}$ curve: increasing with increasing $\rho_0$ or decreasing $\beta$. For $\theta$, larger values lead to an increase in both the duration of the dip and its amplitude. $\varphi$ variations induce changes in the shape of the $\Delta\rm{RM}$ curve; positive $\varphi$ leads to a taller and wider peak and a tiny dip, while a more negative $\varphi$ results in a sharper dip (positive and negative $\varphi$ correspond to the positive and negative projections of $\vec{e}_{\rm{pre}}$ onto the x-axis in Figure \ref{fig:1}, respectively). For $\omega$, it regulates how quickly overall amplitude of the $\Delta\rm{RM}$ curve oscillates over each orbital period. The larger $\omega$, the faster the oscillation. For the LoS, variations in $\lambda_1$ significantly affect proportion and amplitude of peaks and dips within one orbital period. A smaller $\lambda_1$ can result in a wider and deeper dip as well as a narrower and lower peak. For $\lambda_2$, its small perturbations can lead to significant changes in amplitude of the $\Delta\rm{RM}$ curve. As $\lambda_2$ decreases, the amplitude of the $\Delta\rm{RM}$ curve increases --- with the peak part increasing far more significantly than the dip part within an orbital period.

We fit $\Delta\rm{RM}_{\rm{E1}}$, $\Delta\rm{RM}_{\rm{E2}}$ and $\Delta\rm{RM}_{\rm{E3}}$ data with our model. We obtained a set of parameters that are reasonable and well fit the RM variation data: $\varepsilon=0.69$, $\rho_0=3.49\times 10^{-15}\,\rm{g\,cm^{-3}}$, $\beta=4.09$, $\theta=0.07\,\rm{rad}$ ($\sim4^\circ$), $\varphi=-0.7\,\rm{rad}$, $\omega=8.01\times 10^{-3}\,\rm{rad\,day^{-1}}$ (corresponding to a precession period of 784.71 days), $\lambda_1=-0.55\,\rm{rad}$, and $\lambda_2-\pi=-9.64\times10^{-2}\,\rm{rad}$ (since $\lambda_2$ is very close to $\pi$, we adopt $\lambda_2-\pi$ to represent its magnitude). The results  are presented in Table \ref{tab:1}, and the fitting curve plotted using the best-fit parameters is shown in Figure \ref{fig:1} as the red curve. 

\begin{table*} 
  \centering
  \caption{Fitting parameters of FRB 20201124A.
  }
  \label{tab:1}
  \begin{tabular}{cc} 
        \hline  
        \hline 
        Parameter & Best fit value \\
        \hline 
        \hline 
        $M_\star\,(M_\odot)$ & 8 \\
        $R_\star\,(R_\odot)$ & 5 \\
        $M_{\rm{NS}}\,(M_\odot)$ & 1.6 \\
        $P_{\rm{orb}}\,\rm{(day)}$ & 73 \\
        $T\,(\rm{K})$ & $2\times10^4$ \\
        $\rm{B}_{0}\,(\rm{G})$ & 10 \\
        $\varepsilon$& 0.69 \\
        $\rho_0\,\rm{(g\,cm^{-3})}$ & $3.49\times 10^{-15}$\\
        $\beta$ & 4.08 \\
        $\varphi\,\rm{(rad)}$ & $-0.7$ \\
        $\theta\,\rm{(rad)}$ & $0.07$ \\
        $\omega\,\rm{(rad\,day^{-1})}$ & $8.01\times 10^{-3}$ \\
        $\lambda_1\,\rm{(rad)}$ & $-0.55$ \\
        $\lambda_2-\pi\,\rm{(rad)}$ & $-9.64\times10^{-2}$ \\
        \hline 
  \end{tabular}
\end{table*}

\section{Discussions and Conclusions} \label{5}

\subsection{Discussions} \label{5.1}
\subsubsection{Other FRBs} \label{5.1.1}
Besides FRB 20201124A, notable RM variations have also been observed in several other FRB sources. 

\begin{figure*}[!t]
    \centering
    \subfigure[FRB 20180301A]{
        \includegraphics[width=0.95\textwidth]{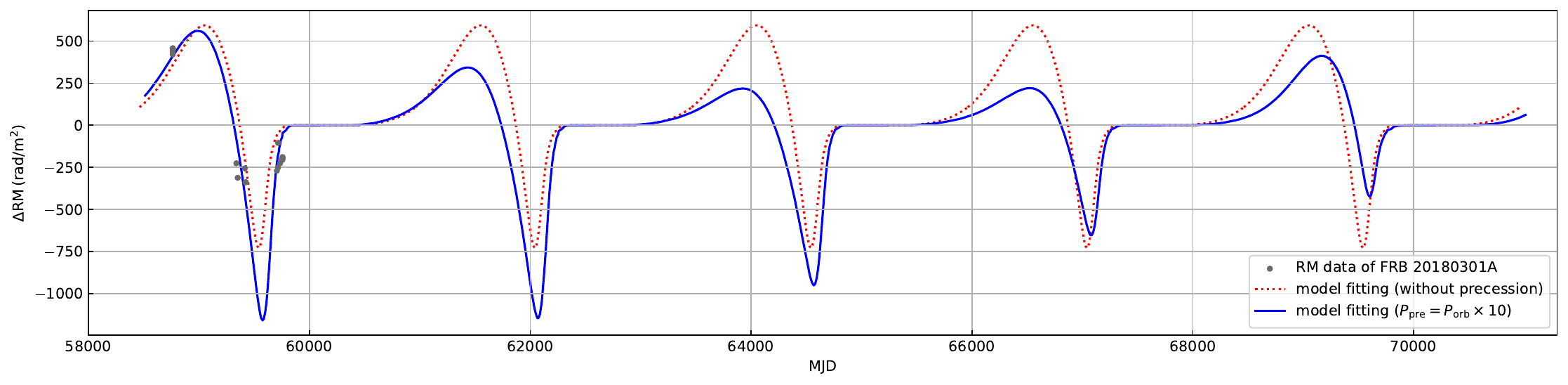}  
        \label{subfig:1}  
    }
    \vspace{0.1cm} 

    \subfigure[FRB 20190520B]{
        \includegraphics[width=0.95\textwidth]{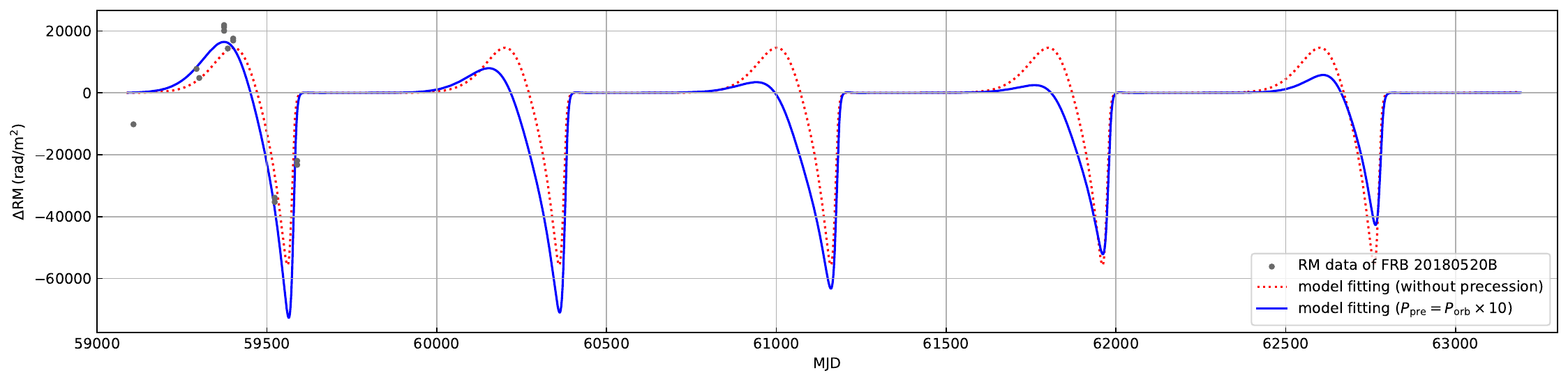}
        \label{subfig:2}
    }
    \vspace{0.1cm}

    \subfigure[FRB 20191106C]{
        \includegraphics[width=0.95\textwidth]{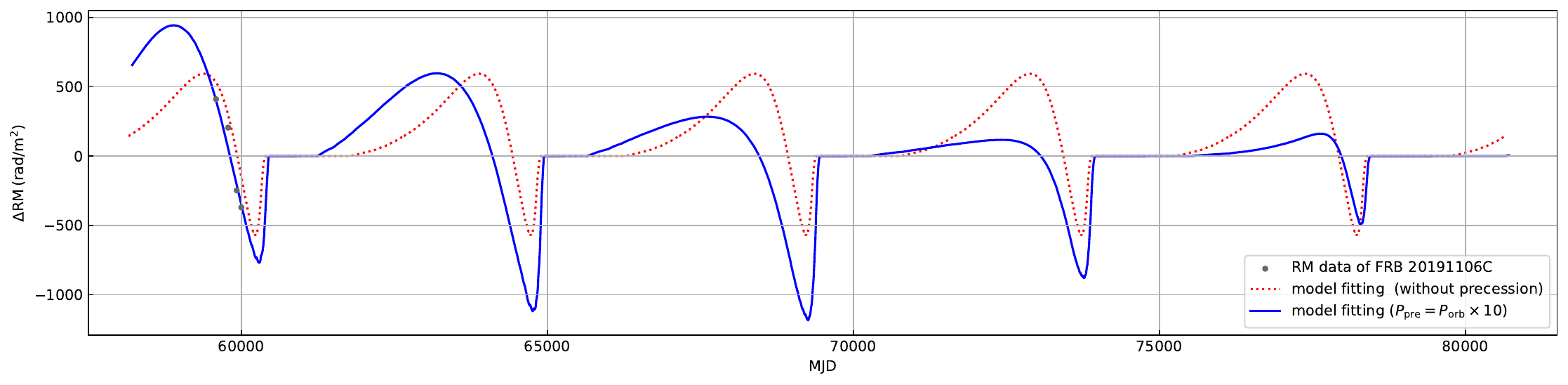}
        \label{subfig:3}
    }
    \vspace{0.1cm}
    
    \caption{Three FRB sources with notable RM variation, each with model fitting RM curves of $P_{\rm{prec}}=10\times P_{\rm{orb}}$ (blue curve) and without precession (dotted red curve). The two curves in each subplot share the same set of parameters except $\theta$ and $\omega$, since these two parameters are both 0 for the no-precession case. Evolutions of these two curves are shown for five orbital periods in each subplot. The observed $\Delta\rm{RM}$s are shown as the gray dots. The parameters of each FRB source are shown in Table \ref{tab:2}.}
    \label{fig:3}  
\end{figure*}

Unlike FRB 20201124A, these FRB sources do not yet have sufficient observational data, making constraint on disk precession loose. We apply our model to three of these FRBs, i.e. FRB 20180301A, FRB 20190520B, and FRB 20191106C, and also offer a prediction for their missing part of RM observations. The $\Delta \rm{RM}$ data of FRB 20180301A, FRB 20190520B, and FRB 20191106C are obtained from \cite{Kumar_2023MNRAS.526.3652K}, \cite{Anna-Thomas_2023Sci...380..599A}, and \cite{Ng_2025ApJ...982..154N}, respectively. The fittings and predictions of these sources are shown in Figure \ref{fig:3}. 

For each source, we present a set of parameters to plot two RM curves: the curve without precession and the curve of $P_{\rm{prec}}=10\times P_{\rm{orb}}$ (the two curves only differ in $\omega$ and $\theta$, since $\omega$ and $\theta$ are set to 0 for the curve without precession). Table \ref{tab:2} lists the parameter sets used for each source.

\subsubsection{Dynamical Analysis of Disk Precession} \label{5.1.2}
In this work, we assume an ideal uniform precession of the Be star with a rigid decretion disk. However, the case would be more complex in practice. In fact, whether a disk can behave as a rigid body depends on whether the interactions between its different parts are sufficiently strong --- such as bending waves, viscous stresses, and self-gravity \citep{Papaloizou_&_Terquem_1995MNRAS.274..987P, Ogilvie_1999MNRAS.304..557O, Ogilvie_2006MNRAS.365..977O, Bate_2000MNRAS.317..773B, Batygin_2012Natur.491..418B, Tremaine_2014MNRAS.441.1408T}. Furthermore, according to \cite{Lai_2014MNRAS.440.3532L}, the circumstellar disk of a massive star in a binary system might undergo precessional evolution, which is driven by the gravitational torques exerted by the companion star tidally. Larger stellar obliquity, i.e., the misalignment between the equatorial plane of the star and the orbital plane, might be induced by such evolution of the disk precession. However, such disk evolution is expected to be significant over a timescale of mega-years, so we do not consider it in our model.

Following the calculation of a rigid disk precession induced by a companion star in \cite{Lai_2014MNRAS.440.3532L}, we estimate the precession period using the binary and disk parameters in Section \ref{4.2}. The disk with the density profile in Equation (\ref{(1)}) can be regarded as a thin disk that spans from the inner radius $R_{\rm{in}}=R_\star$ to $R_{\rm{out}}=\eta R_\star$ with surface density:
\begin{equation}
    \Sigma(r)=\Sigma_0(r/R_\star)^{-\beta+3/2},
\label{(14)}
\end{equation}
\noindent where $\Sigma_0$ is the disk surface density at $R_{in}$. The total disk mass is
\begin{equation}
    M_{\rm d}=\int dM_{\rm d}=2\pi\int_{R_{\rm{in}}}^{R_{\rm{out}}} r\Sigma(r) dr=\frac{4\pi \Sigma_0R_\star^4}{7-2\beta}(\eta^{7/2}-1),
\label{(15)}
\end{equation}
and for the disk angular momentum vector $\vec{L}_{\rm d}=L_{\rm d}\,\vec{e}_{{L}_{\rm d}}$ (where $\vec{e}_{L_{\rm d}}$ is the unit vector) can be calculated as
\begin{equation}
    L_{\rm d}=\int rM_{\rm d}(r)v_k(r)dr=\frac{2\pi\Sigma_0R_\star^{5/2}\sqrt{GM_\star}}{4-\beta}(\eta^{4-\beta}-1).
\label{(16)}
\end{equation}
The gravitational torque exerted on the disk by the external magnetar companion is:
\begin{equation}
\begin{aligned}
        \vec{T}_{\rm d} &=-\int^{R_{\rm{out}}}_{R_{\rm{in}}} dM_{\rm d}\frac{3GM_{\rm{NS}}}{4a_{\rm b}^3}\cos\theta\,\vec{e}_{L_{\rm b}}\times \vec{e}_{L_{\rm d}}\\
        &=\frac{3\pi GM_{\rm{NS}}\Sigma_0R_\star^4\cos\theta}{(11-2\beta)a_{\rm b}^3}(\eta^{11/2-\beta}-1)\,\vec{e}_{L_{\rm b}}\times \vec{e}_{L_{\rm d}}\\
        &=\omega\vec{e}_{L_{\rm b}}\times \vec{L}_{\rm d},
\end{aligned}
\label{(17)}
\end{equation}
\noindent where $\vec{e}_{L_{\rm b}}$ is the unit vector along the binary angular momentum axis, $a_{\rm b}$ is the binary separation, $\theta$ is the inclination angle between $\vec{e}_{L_{\rm b}}$ and $\vec{e}_{L_{\rm d}}$, and $\omega$ is the angular velocity of the disk precession. 
Thus, the precession period of the disk, $P_{\rm{prec}}$, can be calculated as
\begin{equation}
\begin{aligned}
    P_{\rm{prec}}&=2\pi/\omega =2\pi/|\vec{T}_{\rm d}/\vec{L}_{\rm d}|\\
    &=\frac{4\pi{a_{\rm b}}^3}{3M_{\rm{NS}}R_\star^{3/2}\rm{cos}\theta}\sqrt{\frac{M_\star}{G}}(\frac{11-2\beta}{4-\beta})(\frac{\eta^{4-\beta}-1}{\eta^{11/2-\beta}-1}).
\end{aligned}
\label{(18)}
\end{equation}

According to Section \ref{4.2}, $\beta=4$ is reasonable for the case of FRB 20201124A. When $\beta=4$, one has
\begin{equation}
\begin{aligned}
    P_{\rm{prec}}&\simeq \frac{4\pi a_{\rm b}^3}{M_{\rm{NS}}R_\star^{3/2}\rm{cos}\theta}\sqrt{\frac{M_\star}{G}}\frac{\ln(\eta)}{\eta^{3/2}-1}.
\end{aligned}
\label{(19)}
\end{equation}
$\eta$ shows an inverse correlation with the precession period $P_{\rm{prec}}$. We adopt the parameter values from Table \ref{tab:1}, i.e., $\theta=0.07\,\rm{rad}$, $M_{\rm{NS}}=1.6M_\odot$, $M_\star=8M_\odot$, and $R_\star=5R_\odot$. We take the time averaged separation for $a_b$:
\begin{equation}
    \langle a_{\rm b} \rangle_t\ =\frac{1}{P_{\rm{orb}}}\int_0^{P_{\rm{orb}}} a_{\rm b}(t) dt=a(1+e^2/2),
\label{(20)}
\end{equation}
\noindent where $a$ and $P_{\rm{orb}}$ are the semi-major
axes of the binary orbit. Here, we adopt $P_{\rm{orb}}=73 \,\rm{day}$ and $e=0.69$ from Table \ref{tab:1} to calculate $a$ (see Equation (\ref{(5)})) and $ \langle a_b \rangle_t\ $. 

From the above formulae, we calculate that $\eta$ spans the range $[100, 200]$, corresponding to $P_{\rm{prec}}$ in the range $[498, 1220]$ days. The $\sim800$-day period in Section \ref{3} corresponds to $\eta\sim141$. To estimate the impact of $\beta$ on the precession period, we fix $\eta$ to 125 in Equation (\ref{(18)}). A $P_{\rm{prec}}$ in the range $[500, 1200]$ days, corresponding to $\beta$ in the range $[3.71, 4.21]$, and in this calculation $\beta=4$ is not included, as $\beta=4$ is a singularity in Equation (\ref{(18)}) (an increase in $\beta$ leads to a corresponding increase in $P_{\rm{prec}}$). This is comparable to $\beta$ in Table \ref{tab:1}. 

This indicates that, with reasonable values of $\eta$ and $\beta$, the model in Section \ref{3} can adequately account for the expected $\sim785$-day precession period in Section \ref{4.2}.

\subsubsection{Absorption of FRBs by the Be Star Disk} \label{5.1.3}
When an FRB penetrates the decretion disk of a Be star, the absorption process by the disk material may occur \citep{Wang_2022NatCo..13.4382W,BZhang2023}. To make sure that the FRB can reach the observer, it is essential to discuss the absorption process.

\begin{figure*}
    \vspace{1cm}
 	\centering
 	\includegraphics[width=1\textwidth, height=0.8\textheight, keepaspectratio]{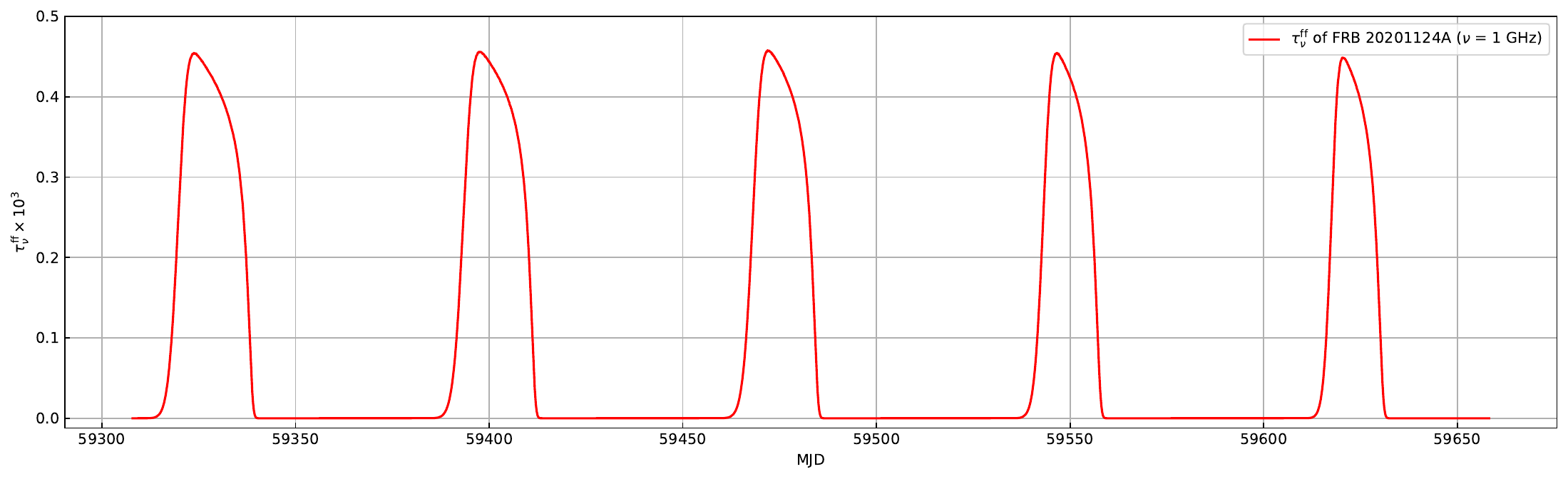}
\caption{The optical depth of free-free absorption contributed by the Be star decretion disk in the case of FRB 20201124A. The evolution time of $\tau^{\rm{ff}}_{\nu}$ consistent with that in Figure \ref{fig:2}, showing the variation of $\tau^{\rm{ff}}_{\nu}$ over 5 orbital periods.} \label{fig:4}
\end{figure*}

The free-free absorption can be important when the medium density is high along the propagation path of the FRB. The optical depth of the free-free absorption at frequency $\nu$, i.e. $\tau^{\rm{ff}}_{\nu}$, can be calculated as follows:
\begin{equation}
\begin{aligned}
    \tau^{\rm{ff}}_{\nu} &= \int_{\rm{LoS}} \alpha_{\nu}^{\rm{ff}} \, dl\\
    &\simeq 8.2\times 10^{-2}\,T^{-1.35}\nu^{-2.1}\int_{\rm{LoS}}n_e^2dl,
\end{aligned}
\label{（21）}
\end{equation}
\noindent where $\alpha_{\nu}^{\rm{ff}}$ is the free-free absorption coefficient, and $\nu$ is the frequency of FRB signal. Here, $T$ and $\nu$ are in units of Kelvin and GHz, respectively. Using the lowest observation frequency of 1 GHz and $T=20,000K$ from Section \ref{4.1}, with parameters of the model from Table \ref{tab:1}, the calculated $\tau^{\rm{ff}}_{\nu}$ is below $5\times 10^{-4}$ (see Figure \ref{fig:4}), which is much less than 1. Thus, in the case of FRB 20201124A, the disk has no effect on the detectability of the radio signal.  

For the other three FRBs in Section \ref{5.1.1}, we also calculate each of their $\tau^{\rm{ff}}_{\nu}$ using the parameters in Table \ref{tab:2}. The observation frequency ranges of FRB 20180301A, FRB 20180520B, and FRB 20191106C are $\sim$ 0.7 to 1.5 GHz, $\sim$ 0.7 to 8 GHz, and $\sim$ 0.4 to 0.8 GHz, respectively \citep{Kumar_2023MNRAS.526.3652K, Anna-Thomas_2023Sci...380..599A, Ng_2025ApJ...982..154N}. Since $\tau^{\rm{ff}}_{\nu}\propto \nu^{-2.1}$, we adopt 0.7 GHz in the calculation of FRB 20180301A and FRB 20180520B, and 0.4 GHz for FRB 20191106C. It turns out that their free-free absorption optical depths are far less than unity.

The synchrotron absorption may also play a role. However, given that our work assumes scenarios after the FRB has been emitted from the emitter environment, we do not consider absorption effects associated with the emitter environment here. On the other hand, since we adopt a cold disk (20,000 K) in our work that lacks relativistic electrons, we can neglect the synchrotron absorption contributed by the disk as the FRB propagates through it.

\subsubsection{The magnetic field in the stellar outflow} \label{5.1.4}

For an early-type massive star, the magnetic field is dipolar within a certain region that extends from the stellar surface \citep{Usov_&_Melrose_1992ApJ...395..575U}. In this region, the stellar outflow moves along the dipolar field and is confined by these closed magnetic field lines near the equatorial plane. This region is defined by the Alfv\'enic surface, where the ram pressure of the outflow is equal to the characteristic value of the magnetic stress components \citep{Mestel_&_Spruit_1987MNRAS.226...57M, Usov_&_Melrose_1992ApJ...395..575U}:
\begin{equation}
    \frac{1}{2}\rho(R_{\rm A})v_{\rm{w}}^2=\frac{\mathrm{B}_{\rm ms}(R_{\rm A})^2}{8\pi},
\label{(22)}
\end{equation}
\noindent where $\rho(r)=\rho_0(r/R_\star)^{-\beta}$ is the density profile of the outflow disk in the equatorial plane according to Equation (\ref{(1)}), $R_{\rm A}$ is the radius of the Alfv\'enic surface (i.e., the Alfv\'en radius), and $v_{\rm{w}}$ is the wind velocity. The wind velocity $v_{\rm{w}}$ can be estimated as the terminal speed of the wind $v_{\rm{w,\infty}}$. $v_{\rm w,\infty}$ of $\sim10^7-10^8$ cm/s is typical for B-type stars from observations \citep{Snow_1981ApJ...251..139S,Krticka_2014A&A...564A..70K}. $\mathrm{B}_{\rm ms}(r)$ denotes the magnetic field strength of the massive star in the out-flowing gas \citep{Usov_&_Melrose_1992ApJ...395..575U, Melatos_&_Johnston_&_Melrose_1995MNRAS.275..381M, Zhao_&_Zhang_&_Wang_&_Dai_2023ApJ...942..102Z}:
\begin{equation}
    \mathrm{B}_{\rm ms}(r)=
    \begin{cases}
        \mathrm{B}_{\rm s}(\frac{r}{R_\star})^{-3}, & (\mathrm{dipolar},\,R_\star < r < R_{\rm A}) \\
        \,\\
        \mathrm{B}_{\rm s} (\frac{R_{\rm A}}{R_\star})^{-3} (\frac{r}{R_{\rm A}})^{-2}, & (\mathrm{radial},\, R_{\rm A}< r < \frac{v_{\rm{w,\infty}}}{v_{\rm{rot}}}R_\star)\\
        \,\\
        \mathrm{B}_{\rm s} (\frac{R_{\rm A}}{R_\star})^{-3} (\frac{r}{R_{\rm A}})^{-1}, & (\mathrm{toroidal},\,r >\frac{v_{\rm{w,\infty}}}{v_{\rm{rot}}}R_\star)
    \end{cases},
\label{(23)}
\end{equation}
\noindent where $\mathrm{B}_{\rm s}$ is the magnetic field strength at the stellar surface, and $v_{\rm{rot}}$ denotes the surface rotation velocity of Be star. For typical values of surface rotation velocity for O or B-type stars are $v_{\rm{rot}}\sim0.15\, v_{\rm{w,\infty}}$. As indicated in Equation (\ref{(23)}), the magnetic field beyond the Alfv\'en surface becomes radial due to the drag effect of the stellar wind. Eventually, the magnetic field changes from radial to toroidal at a larger distance, which is consistent with Equation (\ref{(4)}). Thus, $R_{\rm A}$ can be obtained from Equations (\ref{(22)}) and (\ref{(23)}).   

For FRB 20201124A, taking into account that the toroidal magnetic field in Equation (\ref{(4)}) and \ref{(23)} is the same, we calculate a surface magnetic field strength $\mathrm{B}_{\rm s} =40\,\rm{G}$, with Alfv\'en radius $R_{\rm A}\sim 2 R_\star\,(10\,R_\odot)$ ($v_{\rm{w}}=10^8$ cm/s is adopted in calculation). In comparison, using the LoS and orbital parameters of FRB 20201124A from Table \ref{tab:1}, the perpendicular distance between the FRB’s LoS and the Be star, $D_\perp$, is calculated to be about $45\sim250\,R_\odot$, which is much larger than $R_{\rm A}$, and also exceeds the range dominated by the radial magnetic field (the calculation results can be seen in Table \ref{tab:3}. Thus, the dipolar magnetic field of the Be star does not contribute to the RM of FRB 20201124A, and the magnetic field along the LoS of FRB 20201124A can be regarded as toroidal. 

For other FRB sources in Section \ref{5.1.1}, we adopt the parameters from Table \ref{tab:2} and $v_{\rm{w}}$ of $\sim10^7-10^8$ cm/s to calculate $\mathrm{B}_{\rm ms}(r)$, $R_{\rm A}$ and the minimum $D_{\rm{\perp}}$ for each case (also see Table \ref{tab:3}). Using a comparison of $R_{\rm A}$ of these sources to $D_{\rm{\perp,min}}$, we find that the FRB signals travel through the disk at locations well beyond $R_{\rm A}$ and in the toroidal magnetic field-dominated regime for each source. Therefore, in this work only the toroidal magnetic field of Equation (\ref{(4)}) is considered. 

The disk may also has a dipolar-like magnetic field of its own according to some researches \citep{Balbus_&_Hawley_1991ApJ...376..214B,Dobler_&_Brandenburg_&_Shukurov_1999ptep.proc..347D}. However, given that the dipole magnetic field strength $\rm{B}_{\rm{d}}$ decays as $r^{-3}$, and the toroidal magnetic field strength $\mathrm{B}_{\mathrm{t}} \propto r^{-1}$, $\rm{B}_{\rm{d}}$ can be neglected compared to $\rm{B}_{\rm{t}}$ at the distance of $r\sim10-10^2\,R_\odot$ (e.g., in the case of FRB 20201124A, using that $\rm{B}_0=10\,G$, $\rm{B}_{\rm{t}}$ is about 1 G at $D_{\rm{\perp,min}}=48\,R_\odot$, while $\rm{B}_{\rm{d}}$ is only about 0.01 G at the same distance).

\subsection{Conclusions} \label{5.2}
In this work, we demonstrate that a Be star-magnetar binary system with a decretion stellar disk is suitable for explaining the peculiar RM variations observed by FAST from MJD 59307 to MJD 59658. 

The RM in the first epoch exhibits a pattern of rising, a deeper drop, rising again, and then entering a steady phase. Such a pattern strongly supports that FRB 20201124A originates from a binary system consisting of a magnetar and a massive star with a decretion disk. In the subsequent two observation epochs, the RM fluctuations decrease significantly compared with $\Delta\rm{RM_{E1}}$. The precession of the Be star along with its decretion disk is commonly used to explain the observed superorbital brightness oscillations of Be/X-ray binary system. 

We propose that the long-term decrease in the observed RM variations can be interpreted by the superorbital modulation from a precessing stellar disk in a Be star-magnetar binary. By fitting the data, we obtain the best parameters that fit the RM curves. The best parameters are presented in Section \ref{4}. 

In addition, we use this model to explain a few other FRBs with similar dramatic RM variation, which can also be well explained and predicted. However, unlike FRB 20201124A, due to insufficient data sampling, the constraint of the model parameters remains limited.

With future continuous follow-up of these FRB sources with interesting RM variations, as well as the discovery of more new FRB sources with similar observational features, the understanding of the origin of such sources will be further advanced. These subsequent observations are also expected to enable better constraints on the model proposed in this paper.

\begin{acknowledgments}
We thank the anonymous referee for detailed and very constructive suggestions that have allowed us to improve our
manuscript. We appreciate Dr. A. M. Chen for useful discussion of the precession. This work is supported by the National Natural Science Foundation of China under grants 12041306, 12473012, 12533005, 12173014 12494575, 12273009 and 12573044, and the National Key R\&D Program of China (Nos. 2023YFC2205901, 2020YFC2201400, SQ2023YFC220007). W.H.Lei. acknowledges support from science research grants from the China Manned Space Project with NO.CMS-CSST-2021-B11.
\end{acknowledgments}

\bibliography{ms}{}
\bibliographystyle{aasjournalv7}

\begin{table*} 
  \centering
  \caption{Parameters of fitting curves for the three FRB sources in Figure \ref{3}}
  \label{tab:2}
  \begin{tabular}{cccc}
    \hline
    \hline
    Parameters & FRB 20180301A & FRB 20190520B & FRB 20191106C \\
    \hline
    \hline
    $M_\star\,(M_\odot)$ & 8 & 8 & 8  \\
    $R_\star\,(R_\odot)$ & 5 & 5 & 5  \\
    $M_{\rm{NS}}\,(M_\odot)$ & 1.6 & 1.6 & 1.6  \\
    $T\,(\rm{K})$ & $2\times10^4$ & $2\times10^4$ & $2\times10^4$  \\
    $\mathbf{\rm{B_0}\,(\rm{G})}$ & 80 & 70 & 80  \\
    $\varepsilon$ & 0.5 & 0.67 & 0.7  \\
    $P_{\rm{orb}}\,\rm{(day)}$ & 2500 & 800 & 4500 \\
    $\rho_0\,\rm{(g\,cm^{-3})}$ & $1.7\times 10^{-12}$ & $1.2 \times 10^{-13}$ & $4.0\times 10^{-13}$  \\
    $\beta$ & 3.5 & 2.8 & 3.4 \\
    $\varphi\,\rm{(rad)}$ & -0.785 & -0.785 & -0.780  \\
    $\theta\,\rm{(rad)}$ & 0.1 & 0.06 & 0.05  \\
    $\omega\,\rm{(rad\,day^{-1})}$ & $1.26\times 10^{-3}$ & $7.85\times 10^{-4}$ & $6.98\times 10^{-4}$ \\
    $\lambda_1\,\rm{(rad)}$ & -0.61 & -0.26 & -0.35 \\
    $\lambda_2\,\rm{(rad)}$ & 3 & 2.9 & 2.6 \\
    \hline
  \end{tabular}
\end{table*}

\begin{table*} 
  \centering
  \caption{The $v_{\rm{w}}$ adopted and the calculation results of $\mathrm{B}_{\rm s}$, $R_{\rm A}$, and $D_{\rm{\perp,min}}$ of each FRB source in Section \ref{5.1.4}}
  \label{tab:3}
  \begin{tabular}{ccccc}
    \hline
    \hline
    Parameters & FRB 20201124A & FRB 20180301A & FRB 20190520B & FRB 20191106C \\
    \hline
    \hline
    $v_{\rm{w}}\,(\rm{cm/s})$ & $1\times10^8$ & $2\times10^7$ & $6\times10^7$ & $5\times10^7$  \\
    $\mathrm{B}_{\rm s}\,(\rm{G})$ & 42 & 117 & 90 & 210  \\
    $R_{\rm A}\,(R_\odot)$ & 10 & 6 & 5.6 & 8  \\
    $D_{\rm{\perp,min}}\,(R_\odot)$ & 48 & 770 & 226 & 583  \\
    \hline
  \end{tabular}
\end{table*}

\end{document}